\documentclass[preprint,11pt]{elsarticle}

\usepackage{amssymb}

\usepackage{amsmath}
\usepackage{mathrsfs}

\journal{}

\begin{document}

\begin{frontmatter}

\newtheorem{remark}{Remark}



\title{Modeling self-organized systems interacting with few individuals: from microscopic to macroscopic dynamics}

 \author[ad]{G. Albi}
 \author[ad]{L. Pareschi}
  \address[ad]{University of Ferrara, Department of Mathematics, Via Machiavelli 35, I-44121 Ferrara, Italy.}
  
\begin{abstract}
In nature self-organized systems as flock of birds, school of fishes or herd of sheeps have to deal with the presence of external agents such as predators or leaders which modify their internal dynamic. Such situations take into account a large number of individuals with their own social behavior which interact with a few number of other individuals acting as external point source forces. Starting from the microscopic description we derive the kinetic model through a mean-field limit and finally the macroscopic system through a suitable hydrodynamic limit.
\end{abstract}
\end{frontmatter}

\section{Introduction}
The aim of this paper is to present different level of descriptions for the dynamic of a large group of agents influenced by a small number of external agent. In a biological context this corresponds to the behavior of a flock or a school of fishes attacked by one or more predators, or the movement of a heard of sheep guided by a sheepdog. Recently such dynamics have been studied also in robotic research, where engineers tried to control the action of a school of fishes introducing a fishbot recognized as leader, see~\cite{robot}.

From the modeling viewpoint this turns out in considering a microscopic dynamic described by classical flocking models like \emph{Cuker-Smale} and \emph{D'Orsogna-Bertozzi et al.} (see~\cite{MR2324245, d2006self}) interacting with a set of few individuals characterized in an analogous way as in~\cite{Colombo}. Moreover, motivated by the works of~\cite{ballerini2008interaction}, we endowed the classical dynamic of interaction both with a \emph{metric} as well as a \emph{topological interaction rule}.

Following the approach in~\cite{MR2744704} we start from the microscopic dynamic, given by a ODEs system, and we
derive two other different level of description: the mesoscopic (or kinetic) level through a \emph{mean-field limit} and the macroscopic level through a suitable \emph{hydrodynamic limit}. At difference to the first order macroscopic models proposed in~\cite{Colombo} here we obtain second order models for the corresponding continuum dynamic.

Finally we report some numerical examples for the solution of the mean field model in a series of test cases. The simulations have been performed using the fast algorithm recently presented in~\cite{BinaryAlg}. Rigorous mathematical results concerning the asymptotic limits just described can be found in~\cite{PredPrey}.

\section{Microscopic model}
We are interested in the study of a dynamical system composed of $N$ individuals and $N^p$ external agents with the following general structure

    \begin{equation}
    \small
        \left\{
        \begin{array}{l}
        \dot{x}_i=v_i  \qquad\qquad\qquad\qquad\qquad\qquad\qquad\qquad i=1,\ldots, N
        \\\\
        \displaystyle \dot{v}_{i}=\frac{1}{N^*}\sum_{j\in\Sigma_{N}^*(x_i)}  
      F(x_i,v_i,x_j,v_j)+\frac{1}{N_p}\sum_{k=1} ^{N_{p}}F^p(x_i,v_i, p_k)
        \\\\
       \dot{p}_h=\varphi_h(t,\textbf{p},\mathcal{A}\rho^N(p_h)) \qquad\qquad\qquad h=1,\ldots, N_p
        \end{array}\right.
    \label{main_micr}
    \end{equation}
    \normalsize
where $(\textbf{x},\textbf{v})_i=(x_i,v_i)$ lives in $\mathbb{R}^{2d}$, $d\geq1$, $i=1,\ldots, N$ and $(\textbf{p})_h=p_h\in\mathbb{R}^{n d}$, with $n=1,2$, $h=1,\ldots, N_p$, and $N_p\ll N$.

Function $F$  describes the interactions inside the swarm, and $F^p$ depicts the interaction with each external agent $p_h$.  
According to the \emph{three zone model}~\cite{Aoki_1982}, $F$ can be decomposed 
in
\begin{equation}
F(x_i,v_i,x_j,v_j)=H(x_i,x_j)(v_j-v_i)+A(x_i,x_j)+R(x_i,x_j)+S(v_i)v_i.
\label{interaction}
\end{equation}
 $H$ characterize the \emph{alignment} term, $A$ the \emph{attraction}, $R$ the \emph{repulsion} and $S$ represent a \emph{self propulsion-friction} term.
The same decomposition holds for $F^p$ if $p_h=(x_h^p,v_h^p)$, i.e.  $n=2$. In a first order model, $n=1$, similar interaction can be considered, bu for the first term we should speak of \emph{consensus} dynamic rather then \emph{alignment}. 

Moreover we endow the model with a \emph{topological} rule of interaction. Each agent will interact 
only with a fix number of agents of their specie,
\begin{equation*}
\Sigma_{N}^*(x_i)=\left\{\textrm{the } N^* \textrm{ closest neighbors respect to } i \right\}.
\end{equation*}
A topological interaction is motivated by some recent studies, see \cite{ballerini2008interaction, MR2744705}, suggesting that in a flock each individual  modifies its position according to few individuals directly surrounding it, no matter how close or how far away those agents are. If $N^*=N$ each agent interacts with all the others and the \emph{topological interaction}  coincides with the global \emph{metric interaction}~\cite{BinaryAlg,MR2596552}.
Functions $\varphi_{h}:[0,+\infty)\times(\mathbb{R}^{nd})^{N_p}\times\mathbb{R}\longrightarrow\mathbb{R}$, describe the evolution in time of each external individual and they depend on the discrete density $\rho^N$ defined as empirical measure
\begin{equation*}\small \rho^N(x,t)=\frac{1}{N}\sum_{i=1}^N\delta(x-x_i(t)).\end{equation*}  \normalsize
According to \cite{Colombo} we define $\mathcal{A}$ as a convolutional operator $\mathcal{A}\rho^N(t)=\rho^N(t)*\eta$, where $\eta$ is a smooth kernel with compact support.
\subsection{Classical swarming models}
The classical swarming models take in account a global interaction between the agents, that corresponds in our case to choose $N^*=N$ neighbors.

\paragraph{Cucker-Smale model} Describes an \emph{alignment dynamic}, see~\cite{MR2324245} and \cite{MR2596552}, 
\begin{equation}\small
H(x_i,x_j)(v_j-v_i)=H(|x_i-x_j|)(v_j-v_i),
\label{CS}
\end{equation}
 \normalsize
where $H(|x_i-x_j|)$ is a function that measures the strength of the interaction between individuals $i$ and $j$ , and depends on the mutual distance, under the assumption that closer individuals have more influence than the far distance ones, as depicted by
\begin{equation*}\small H(r)=\frac{1}{(1+r^2)^\gamma}, \end{equation*}\normalsize
where  $\gamma\geq0$  discriminates the behavior of the solution, see \cite{MR2324245,MR2596552} and \cite{MR2744704}.

\paragraph{D'Orsogna, Bertozzi et al.}, considers a \emph{self-propelling, attraction and repulsion dynamic}, see~\cite{d2006self} 
\begin{equation}
\small
 A(x_i,x_j)+R(x_i,x_j)+S(v_i)v_i=-\nabla_{x_i}W(\left|x_j-x_i\right|)+(\alpha-\beta|v_i|^2)v_i
  \label{AttRep}
\end{equation} \normalsize
 where $W:\mathbb{R}^d\longrightarrow\mathbb{R}$ is a given potential modeling the short-range repulsion and long-range attraction and $\alpha,\beta$ are positive parameters.
 A possible description is given by the following \emph{power law}
    \begin{equation*}\small
      W(r)=\frac{r^a}{a}-\frac{r^b}{b},
    \label{pot}
    \end{equation*}
     \normalsize 
where $a>b>0$ are positive parameters.

Both the models take in account symmetric interaction between agents, which correspond to the conservation of the first momentum. This assumption isn't really realistic if we want to model group of animals, to improve the models other features  have been introduced as a \emph{perception cone}~\cite{BinaryAlg, MR2596552} or the concept of \emph{relative distance}~\cite{motsch2011new}. In these cases we lose the interaction symmetry and consequently the conservation of the first momentum, note that also the \emph{topological interaction} breaks the symmetry.
 \section{Kinetic model}
 In order to retrieve a mesoscopic description of the system (\ref{main_micr}) we proceed through a \emph{mean-field} limit, as already done in~\cite{ PredPrey, MR2596552, MR2744704}. Since the limit is done only for the first set of equations, which describe the evolution of the swarm, we obtain an hybrid model made by one kinetic equation and the ODEs  system governing the external agents.
 
The basic idea of the \emph{mean-field} limit is to derive, through a weak argument, a single evolutionary equation for $f^N$, the empirical measures defined as 
\begin{equation*}\small
f^N(x,v,t)=\frac{1}{N}\sum_{i=1}^{N}\delta(x-x_i(t))\delta(v-v_i(t)).
\end{equation*}
A second step is to show that the limit $f^N\longrightarrow f$ for $N\longrightarrow\infty$ holds true, then rigorous derivation of the limiting kinetic equation can 
be proved.  A well-posedness theory for this derivation has been developed in~\cite{MR2596552} for a general set of swarming models.

We remark that in the mean-field limit the analogous of the \emph{topological set}, $\Sigma_N^*(x)$, of interaction is described by a characteristic function on the ball $\mathcal{B}(x,R^*)$, with center $x$ and radius $R^*$ such that
\begin{equation*}
\small
R^*=\min\left\{R \left| R>0,\rho^*=\int_{\mathbb{R}^d}\int_{\mathcal{B}(x,R)} f(x,v)dxdv\right. \right\}.
\normalsize
\end{equation*}
The model at mesoscopic level reads 
\begin{equation}\small
 \left\{
        \begin{array}{l}
\partial_t f+ v \cdot\nabla_x f=-\nabla_v\cdot(\mathcal{E}^*(x,v)f)-\nabla_v\cdot (\mathcal{E}^p(x,v)f).
 \\\\
 \dot{p}_h=\varphi_h(t,\textbf{p},\mathcal{A}\rho(p_h)) \qquad\qquad\qquad h=1,\ldots, N_p
        \end{array}\right.
         \label{main_kin}
\end{equation} \normalsize
where $\rho(x,t)=\int_{R^d}f(x,v,t)dv$.

\begin{remark} If we consider the decompositions (\ref{interaction}) in the case of \emph{Cucker-Smale} and \emph{D'Orsogna-Bertozzi et al.} models and $p_h=(x_h^p,v_h^p)$, we have a more particular formulation for $\mathcal{E}^*$ and $\mathcal{E}^p$, in this way
\begin{equation}
\begin{aligned}
\mathcal{E}^*(x,v)=&\frac{1}{\rho^*}\int_{\mathbb{R}^d}\int_{\mathcal{B}(x,R^*)}H(|x-y|)(w-v)f(y,w)dydw+\\&-\frac{1}{\rho^*}\int_{\mathcal{B}(x,R^*)}\nabla_{x}W(|x-y|)\rho(y)dy+S(v)v
\end{aligned}
\end{equation}
and
\begin{align}\small
\mathcal{E}^p(x,v)=\frac{1}{N_p}\sum_{k=1}^{N_p}H^p(|x-x_k^p|)(v_k^p-v)-\frac{1}{N_p}\sum_{k=1}^{N_p}\nabla_{x}W^p(|x-x_k^p|)
\end{align} \normalsize
\end{remark}

\section{Hydrodynamic model}
We are interested also in the macroscopic description of the system, from a numerical point of view this corresponds to reduce the dimensionality of the problem in such a way that simulations become affordable.
As done in ~\cite{PredPrey, MR2744704, MR2425606} we define 
\begin{equation*}
\small\rho:=\rho(x,t)=\int fdv,\quad \rho u:=\rho u(x,t)= \int v fdv,\quad T:=T(x,t)=\int |v-u|^2 fdv,
\end{equation*}
in order to obtain a system of equations which describe the evolution of these quantities, we integrate the kinetic equation in (\ref{main_kin})  against $dv$  and $vdv$.

According to~\cite{MR2744704}  we impose the closure the momentum assuming that the fluctuations are negligible, i.e., that
the temperature $T(x, t) = 0$, and the velocity distribution is monokinetic:  $f(x,v,t)=\rho(x, t)\delta(v-u(x, t))$. 
The previous assumptions lead to the following hydrodynamic system 
\begin{equation}\small
 \left\{
        \begin{array}{l}
\partial_t \rho+div_x(\rho u)=0,
\\\\
\partial_t (\rho u)+\nabla_x\cdot(\rho u \otimes u)=\mathcal{F}^*(x,u)\rho(x,t)+\mathcal{F}^p(x,u)\rho(x,t),
 \\\\
 \dot{p}_h=\varphi_h(t,\textbf{p},\mathcal{A}\rho(p_h)) \qquad\qquad\qquad h=1,\ldots, N_p
        \end{array}\right.
\end{equation} \normalsize
In the particular case of Cucker-Smale and D'Orsogna Bertozzi et al. model and $p_h=(x_h^p,v_h^p)$ the second term reads
\begin{equation}
\begin{aligned}\small
\mathcal{F}^*(x,u)=&\frac{1}{\rho^*}\int_{\mathcal{B}(x,R^*)}H(|x-y|)(u(y,t)-u(x,t))\rho(y,t)dy+\\
-&\frac{1}{\rho^*}\int_{\mathcal{B}(x,R^*)}\nabla_xW(|x-y|)\rho(y,t)dy+S(u)u
\end{aligned}
\end{equation}
\begin{equation}
\begin{aligned}\small
\mathcal{F}^p(x,u)=&\frac{1}{N_p}\sum_{k=1}^{N_p}H^p(|x-x_k^p|)(v_k^p-u(x,t))
-\frac{1}{N_p}\sum_{k=1}^{N_p}\nabla_xW^p(|x-x_k^p|)
\end{aligned} \normalsize
\end{equation}

\section{Numerical examples}
We show here two numerical examples taken from~\cite{Colombo} for two different biological dynamics. In order to solve the kinetic part of the model, the simulations are performed using the Asymptotic Binary Interaction algorithms proposed in~\cite{BinaryAlg}. We recall that the algorithm is based on a stochastic routine which use $N_s$ sample particles. The overall cost is $O(N_s)$.\\ 
The dynamics considered for the swarm in both the cases are
\begin{equation*}\small
F(x_i,v_i,x_j,v_j)=H(|x_i-x_j|)(v_j-v_i)+\nabla_{x_i}W(|x_i-x_j|),\\
\end{equation*} \normalsize
and  only a \emph{repulsion} dynamic respect to $p_h$ in
$F^p$, given by the $W^p(r)=-r^c/c$ with $c>0$, and where $r=|x_i-p_h|$.
\paragraph{Shepherd dogs} 
\begin{figure}[t]
\begin{center}
\includegraphics[scale=0.27]{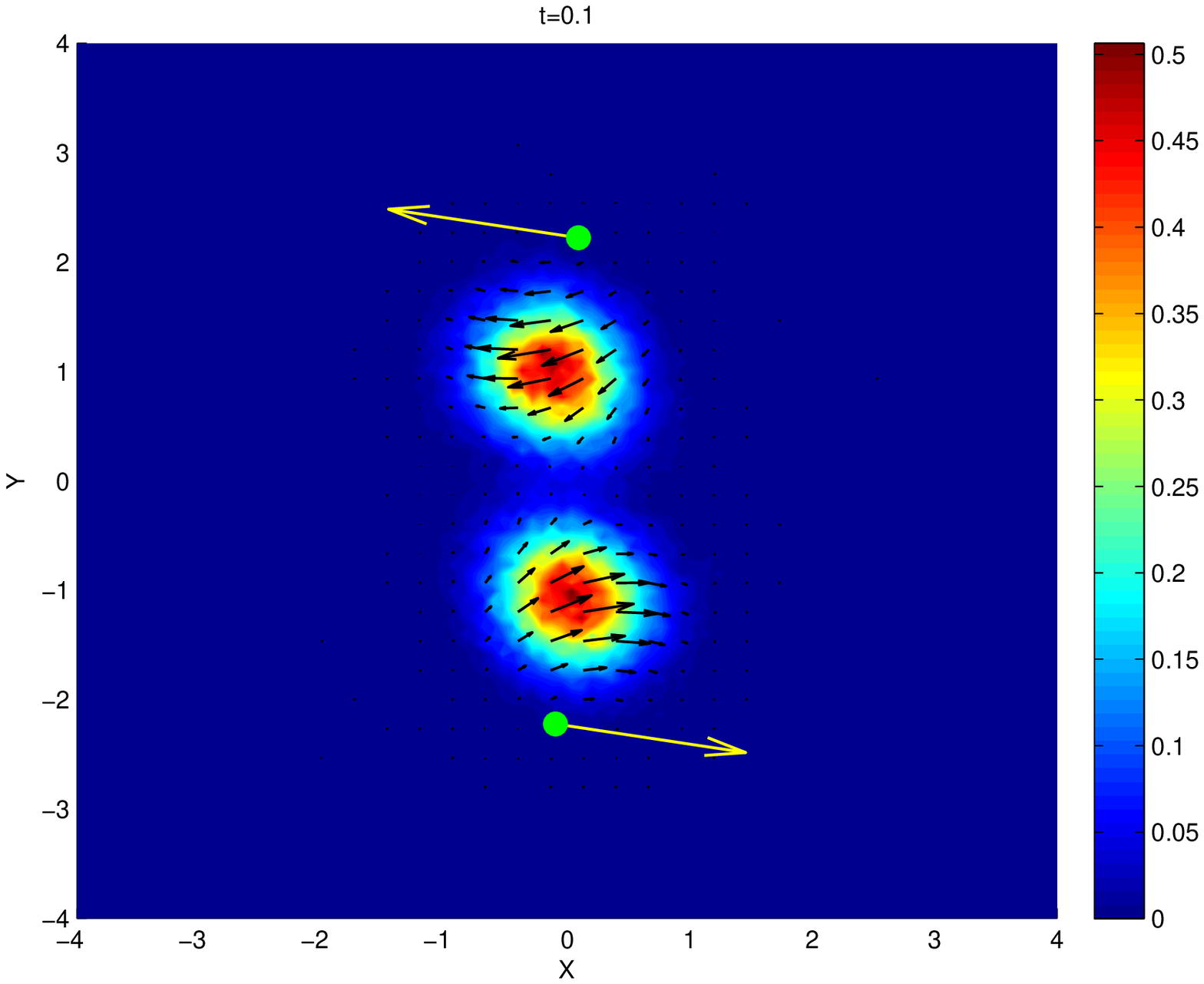}
\includegraphics[scale=0.27]{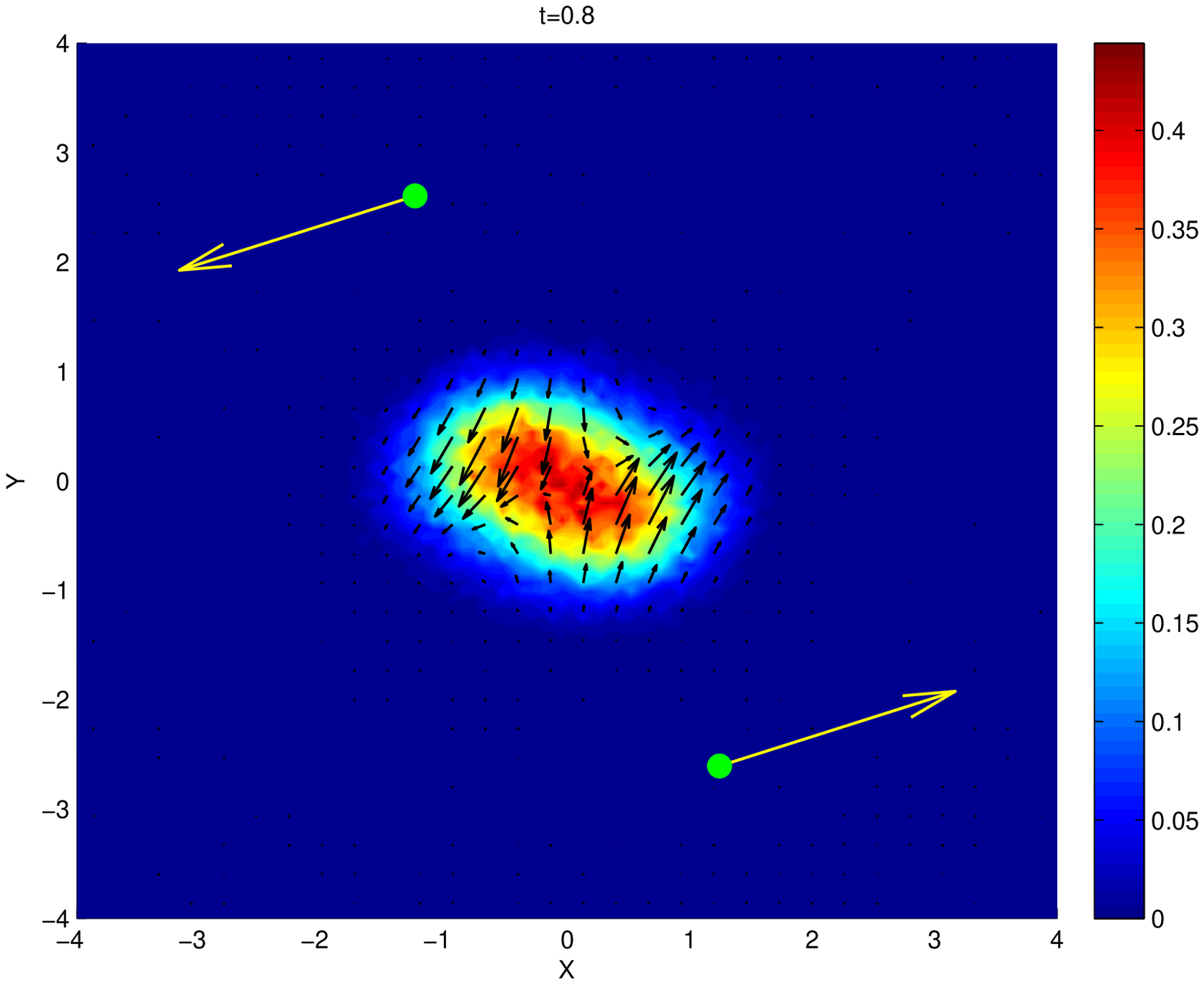}
\includegraphics[scale=0.27]{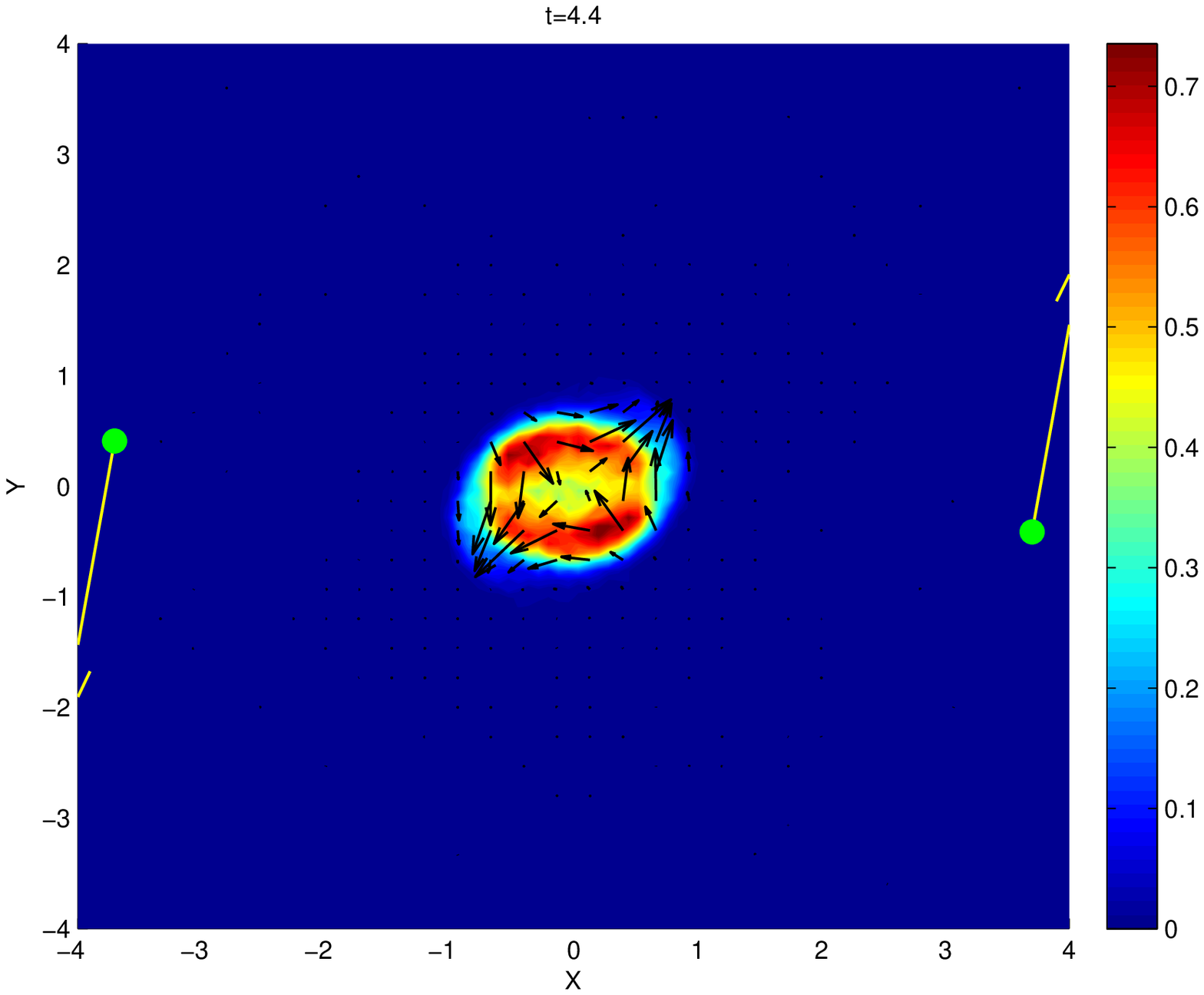}
\end{center}
\caption{
Shepherd dogs. With parameters $a=2.5$, $b=0.1$, $\gamma=0.45$ for models (\ref{CS}) and (\ref{AttRep}).  }
\end{figure}
The simulation represents the evolution of a swarm controlled by two leaders $p_1, p_2\in\mathbb{R}^{2n}$, $n=1$, interacting with the swarm in the following way 
\begin{align*}\small
&\varphi_{h}(t,\textbf{p},s_h)=V_p\frac{s_h^\perp}{\sqrt{1+|s_h|^2}}; \quad V_p=300,\quad r_p=5,\quad h=1,2.
\\
&\eta(x)=\frac{3}{\pi r_p^6}(\max\{0,r_p^2-|x|^2\})^2, \quad s_h:=\mathcal{A}\rho(x^p_h)=(\rho*_x\nabla\eta)(x^p_h).
\end{align*} \normalsize
\paragraph{Swarm attacked by predator}
\begin{figure}[t]
\begin{center}
\includegraphics[scale=0.27]{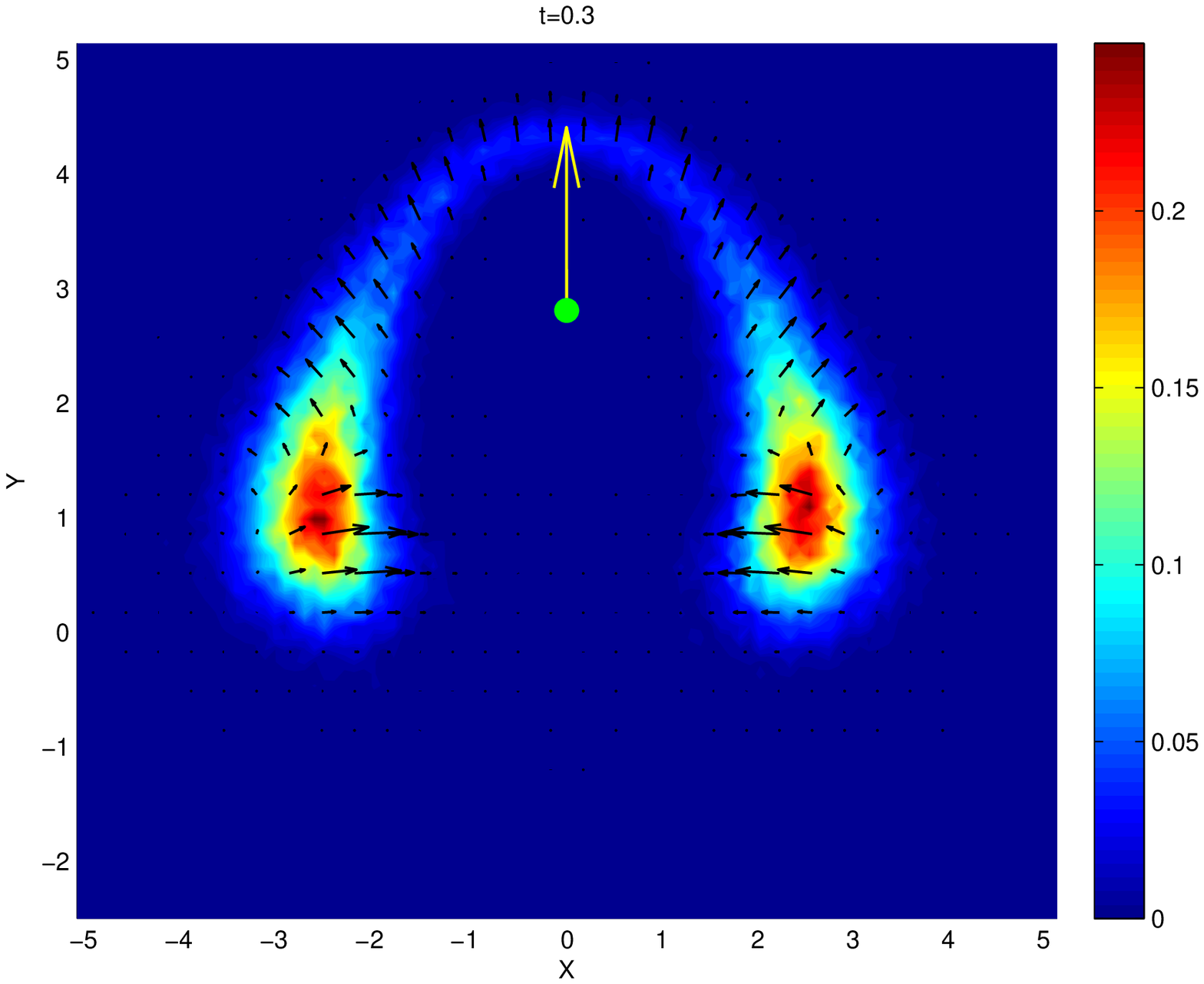}
\includegraphics[scale=0.27]{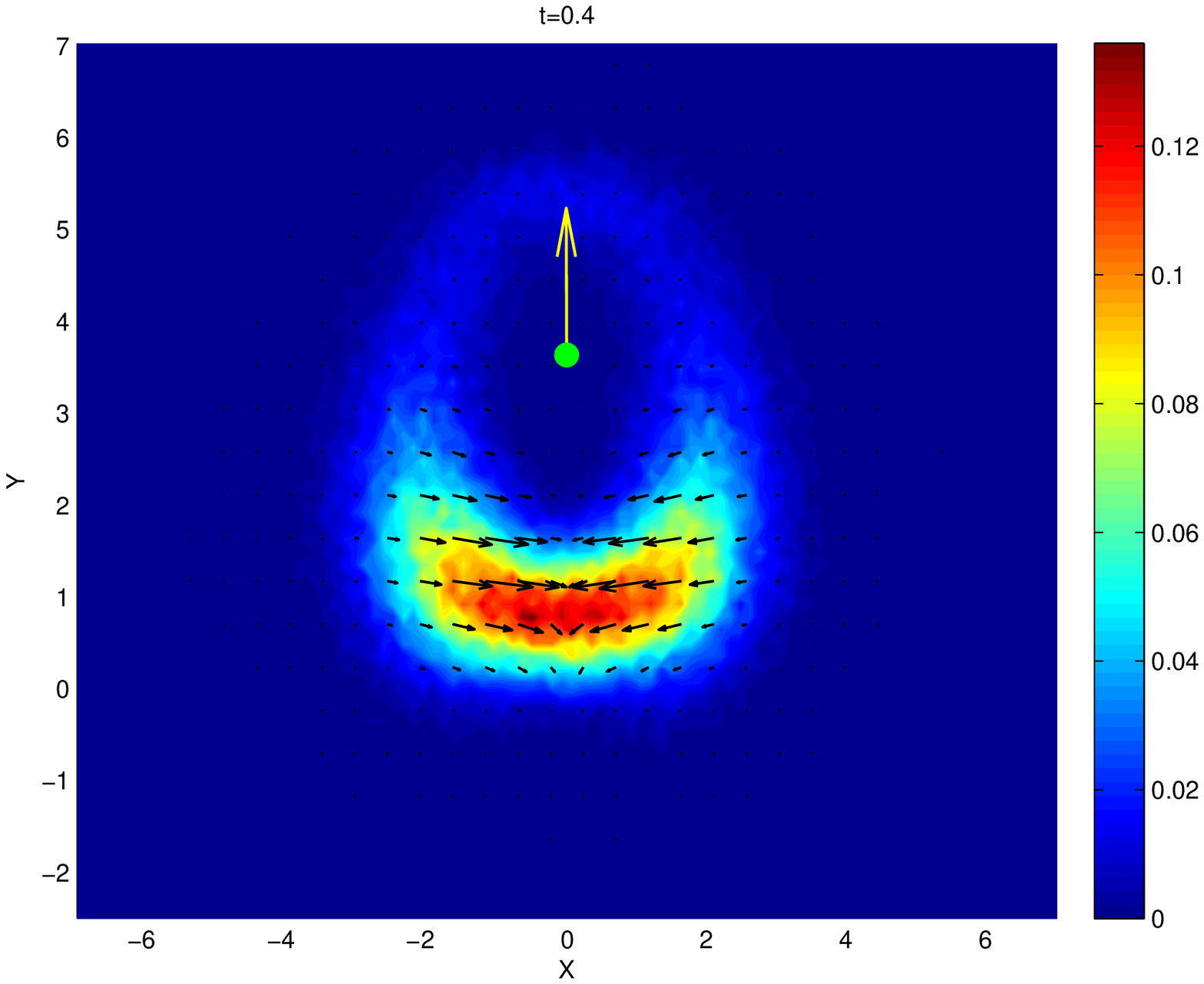}
\includegraphics[scale=0.27]{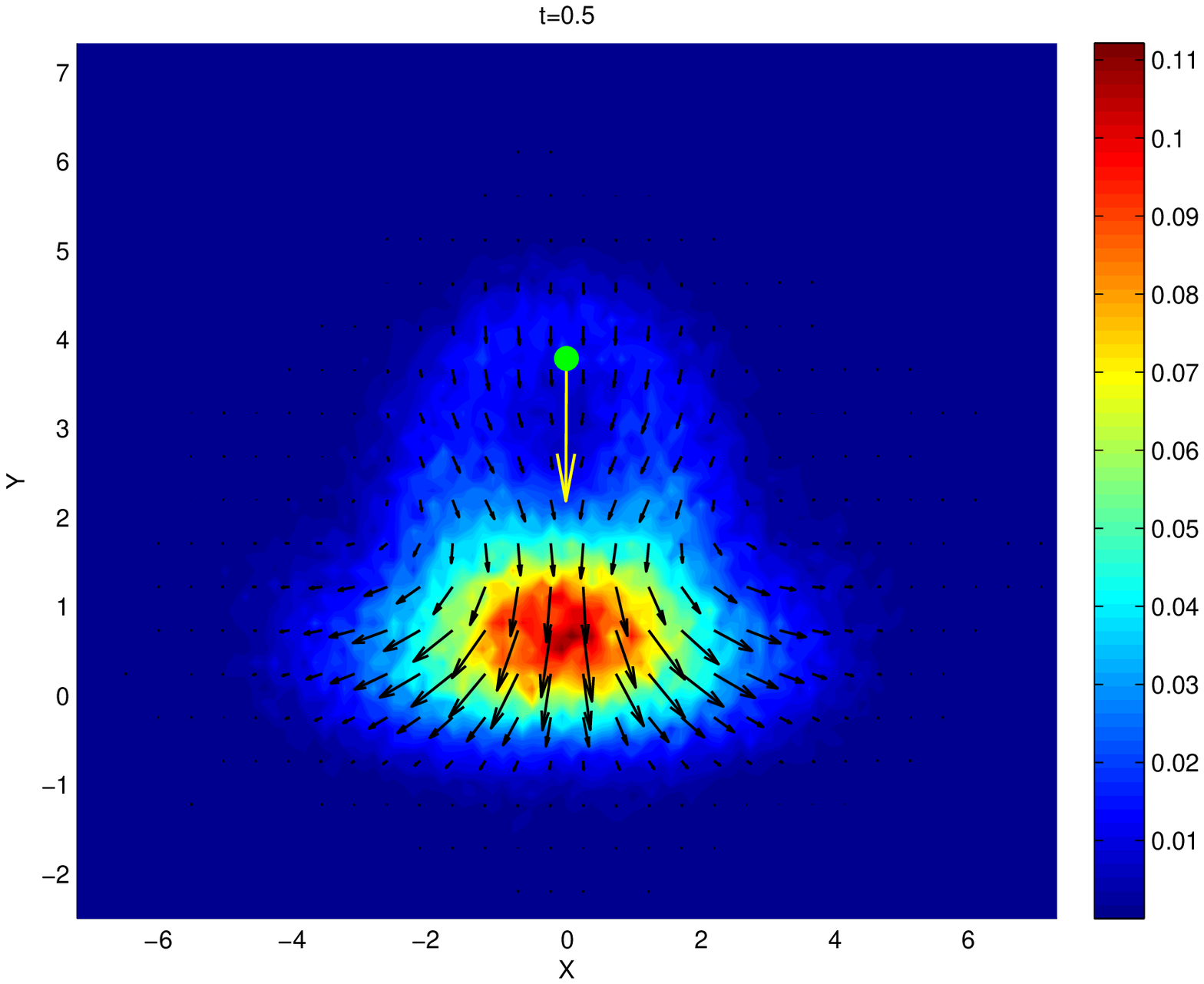}
\end{center}
\caption{Swarm attacked by a predator.  With parameters $a=4$, $b=2$, $\gamma=0.45$ for models (\ref{CS}) and (\ref{AttRep}). } 
\end{figure}
We consider the evolution of a swarm which undergoes the action of a predator. The predator is modeled by the evolution of $p=(x^p,v^p)\in\mathbb{R}^{2n}$, $n=2$ and its evolution is lead by the following potential
\begin{align*}\small
&\varphi(t,p,s)=(v^p, V_p s);\quad V_p=1500,\quad r_p=5.
\\
&\eta(x)=\frac{3}{\pi r_p^6}(\max\{0,r_p^2-|x|^2\})^2, \quad s:=\mathcal{A}\rho=\rho*\nabla\eta.
\end{align*} \normalsize

\bibliographystyle{elsarticle-num}
\bibliography{<your-bib-database>}



\end{document}